# Vacuum Controls and Diagnostics

*G. Pigny, A. Rocha*
CERN, Geneva, Switzerland

**Abstract**

This paper describes the CERN's vacuum control system from the field devices to the Supervisory Control and Data Acquisition software.

First, a particular attention is given to the environment present in the accelerators, like noise coupling and ionizing radiation, which can affect the quality of the measurements and the reliability of the system.

Then, the main vacuum instruments and their associated conditioning circuits and controllers are presented, before to introduce the hardware interlock logic and alarms used for the vacuum system and the machine protection.

Finally, the Supervisory Control and Data Acquisition software and its architecture are described, including data engineering and the main functionalities provided to the users for controls and diagnostics.

**Keywords**

PLC, SCADA, controller, Profibus, gauge, pump, valve, radiation, noise, data, server, network.

## 1    Introduction

Figure 1 shows the three layers of the architecture of the vacuum control system: the filed layer, the control layer and the supervision layer [1].

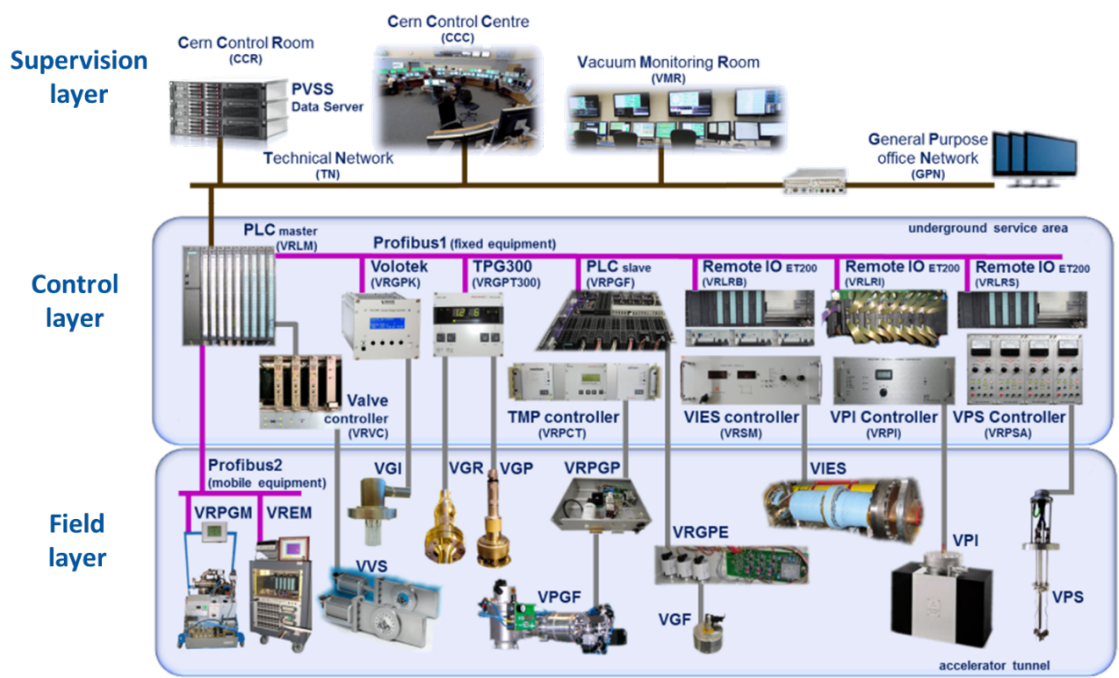

**Figure 1 - Vacuum controls architecture**

The **field layer** concerns all vacuum instrumentation installed on the beam pipe or nearby, usually exposed to ionizing radiation. There are vacuum sector valves, different types of gauges, fixed pumping groups, ion and sublimation pumps, some electronics and all the cables connected to them. In accelerator areas where the radiation dose is considered as acceptably low, and where the distances to service areas are quite large, radiation tolerant electronics are installed directly in the tunnel. In one hand, it allows to cope with measurement or control issues due to the length of cables (e.g. resistance of the wires, leakage current). On the other hand, it allows a significant reduction of cabling costs. A dedicated dynamically-configured Profibus (a serial communication link for field equipment) network connects a PLC (Programmable Logic Controller), installed in service areas, to the mobile equipment, installed in the tunnel when the machine is shut-down. Examples are mobile pumping groups and mobile bake-out stations.

The **control layer** is based on PLCs, controllers and power supplies. These are kept in service areas, away from the accelerator tunnels to prevent radiation damage. While small accelerators or installations are controlled by a single PLC, wider machines have one PLC at each underground service area. A Master PLC accesses the field equipment (i.e. gauges, pumps, valves) through controllers or power supplies. When equipped with the corresponding interface, they can communicate with PLCs directly via Profibus; this minimizes the complexity and price of cabling and allows for a wider exchange of information and configuration parameters. This is the case of the controllers for Bayard-Alpert gauges, for the Pirani and Penning gauges, and for recent DC power supplies. In addition, the fixed pumping groups and their turbo-molecular pump controllers are managed by a dedicated Slave PLC, connected to the Master PLC by Profibus. Controllers for sector valves are directly connected to individual I/O (Input/Output) channels of the Master PLC; power supplies for Sputter ion pumps are connected to remote-IO stations (Input-Output modules connected to the Master PLC via Profibus).

The **supervision layer** is based on a Supervisory Control And Data Acquisition (SCADA) architecture. PLCs exchange data with the SCADA Data-Server, where incoming data is archived, processed, and displayed to the end user via dedicated consoles. The PLCs, control rooms and the SCADA Data-Server communicate through Ethernet in a protected and restricted Technical Network (TN); users can access the vacuum SCADA applications from the General Purpose Network (GPN) via special machines called terminal servers. Independently of the number of PLCs used, there is only one SCADA Data-Server per accelerator complex.

## 2 Environment

### 2.1 Electromagnetic Compatibility (EMC)

Vacuum control systems are distributed systems and have multiple enclosures (cabinets, racks equipment frame, etc.) that are physically separated in different rooms, and buildings. There are also a multiple interconnection I/O cables, usually long, between individual elements of the system. The elements of the system are fed power from different sources, for example different branch circuits, if within a single building, or even different transformer banks if the elements are located in different buildings. In this situation, ground loops may be a problem and source of noise and interference. This is especially true when multiple ground points are separated by large distances and are connected to the ac power ground, or when low-level analog circuits are used. In these cases, it may be necessary to provide some form of discrimination or isolation against the ground path noise.

There is no single solution to the noise-coupling problem. As shown in Figure 2, there are four principal noise "pick up" or coupling mechanisms: conductive, capacitive, inductive, and radiative. Conductive coupling results from sharing currents from different circuits in a common impedance. Capacitive coupling results from time-varying electric fields in the vicinity of the signal path. Inductive or magnetically coupled noise results from time-varying magnetic fields in the area enclosed by the signal circuit. If the electromagnetic field source is far from the signal circuit, the electric and magnetic field coupling are considered combined electromagnetic or radiative coupling [2].

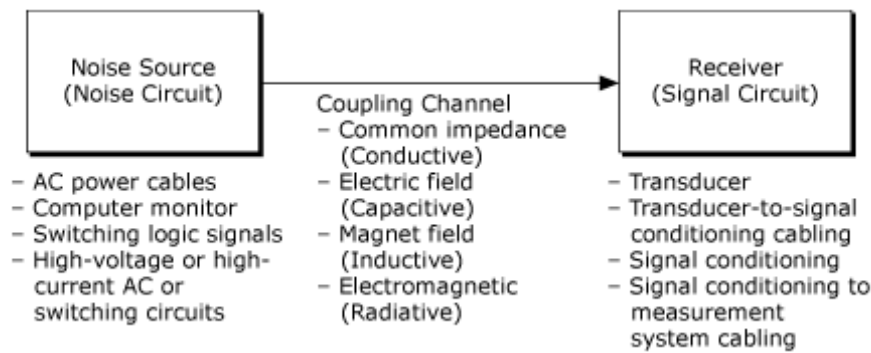

**Figure 2 – Noise-Coupling Problem Block Diagram**

## 2.2 Ionizing radiation environment

Beam interactions with residual gases, collimators or other equipment (e.g. due to beam instabilities) are the source of ionizing radiation. This complex radiation field is composed of different particles at various energies. Electronic components and systems exposed to such radiation fields will experience three different types of radiation damage: Single Event Error (SEE) proportional to High Energy Hadron fluence (HEH), displacement damage proportional to the 1MeV equivalent neutron fluence, and damage from the Total Ionizing Dose (TID) [3] [4].

The majority of controllers that are located in service areas, at the surface, or underground, are shielded from the ionizing radiation generated by the accelerators. In the case of the LHC, as the beam energy and intensity are very high, some of the underground service areas can see a non-negligible HEH. Consequently, the electronics installed in those areas may suffer SEEs and can result in an increase of equipment failures leading to beam dumps, which are time consuming for the machine operation. Several solutions, such as additional shielding, relocation or redundancy of critical controls must be taken into consideration.

All the equipment in the accelerator tunnels are subject to ionizing radiation. Depending on their location, the levels of radiation may vary. For example, in the LHC arcs, the TID is less than 10 Gy/y, whereas in Long Straight Section (LSS) it is higher than 100 Gy/y.

In low radiation zones, some electronics are installed in the tunnel. They are locally powered and send their readings through one or several cables to the controllers in the service areas. To avoid large measurement errors and premature failure of equipment, only radiation tolerant electronics can be installed in the tunnel. For that, active components have to be tested in an irradiation facility (component level) before being used in any radiation tolerant design. Other tests have to be carried-out with several components assembled together (subsystem level) and with the final electronics (system level). This process allows to fully characterize the cards installed in the tunnel and to know their behaviour with time when exposed to radiation.

In high radiation areas, no electronics are installed in the tunnel. Instruments are directly connected to their controllers, installed in service areas. Cables accumulate radiation over the time, which progressively degrade their physical characteristics. In addition, transient effects might be observed in high HEH fluence areas, due to radiation-induced current in HV cables, which can lead to parasitic pressure spikes in measurements.

# 3 Vacuum gauges and signal conditioning

## 3.1 Membrane vacuum gauges (piezoresistive)

Piezoresistive based transducers rely on the piezoresistive effect, which occurs when the electrical resistance of a material changes in response to the applied mechanical strain. When piezoresistors are placed in a Wheatstone bridge configuration and attached to a pressure-sensitive diaphragm, a change in resistance is converted to a voltage output, which is proportional to the applied pressure.

Piezoresistive gauges are used to measure pressures up to several thousands of mbar with an uncertainty of around 10%. The bridge can be supplied by a constant voltage or current source. The latter is preferred since it allows the connection of a long cable between the conditioning circuit and the gauge without any measurement error due to the resistance of the wires.

The first stage of the conditioning circuit is composed by a differential amplifier with a gain $A_D$ to measure the differential voltage $V_D$ coming from the bridge. The differential voltage $V_D$ at the output of the bridge is equal to the excitation current $I_{EX}$ times the resistance variation $\Delta_R$ due to the deformation of the membrane, and thus, to the pressure:

$$V_{OUT} = A_D . V_D = A_D . I_{EX} . \Delta R$$

The next stages are used for gain, offset compensation, and provide filtering before being digitized by an ADC (Analog to Digital Converter). After the conversion, a microcontroller can compute the pressure using the linear relation between voltage and pressure.

## 3.2 Thermal conductivity vacuum gauges (Pirani)

Thermal conductivity vacuum gauges are pressure-measuring instruments for medium and low vacuum that measure the pressure-dependent thermal loss of a heated element, usually a wire, through the gas. Pirani gauges cover pressure range from 1000 to $10^{-4}$ mbar with an uncertainty of around 30%.

A filament is heated and maintained at a constant temperature. The current needed to maintain this temperature depends on the gas conductivity and therefore is a measure of pressure.

The filament constitutes one element of a Wheatstone bridge, which is self-compensated by an operational amplifier in a feedback loop configuration. The voltage output $V_{OUT}$ given by the first stage of the conditioning circuit is a nonlinear function of the pressure, and depends of the geometry of the gauge itself:

$$V_{OUT} = \sqrt{2.R_f.\epsilon.\left(p_0 + \frac{p}{1+g.p}\right)}$$

The constant $\epsilon$ (sensitivity) includes gas characteristics, $g$ is a factor determined by geometry, and $p_0$ corresponds to the offset pressure in the lower limit of the measuring range.

The next stages are used for gain and offset compensation, and provide filtering before being digitized by an Analog to Digital Converter (ADC). After the conversion, a microcontroller computes a non-linear reverse transfer function from voltage to pressure.

## 3.3 Cold cathode ionization gauges (Penning)

The operation principle of these gauges for low pressures consist of using a gas discharge ignited between two metal electrodes (anode, cathode) by applying a sufficiently high DC voltage in the kV range. The gas discharge current is pressure dependent and thus used as a measured quantity. In addition, a magnetic field is used so that the electron paths from the cathode to the anode are stretched considerably by forcing the electrons onto a spiral path. This type of gauge is also called a cross field gauge. A special type of Penning gauge is mainly used nowadays, called inverted magnetron cold cathode gauge, and covers the range from $10^{-5}$ down to $10^{-11}$ mbar with an uncertainty of around 50%. Within a broad range, the ionization current $I^+$ is a measure of the pressure p:

$$I^+ = K.p^m$$

The exponent $m$ depends on the precise design of the gauge and is in the range $m$ = 1-1.4.

In the low range of measurement, the current can be as low as pA; a high quality HV triaxial cable is used for signal transmission. The conditioning circuit is composed by an HV DC power supply paired with a logarithmic ammeter (see chapter 3.5) to measure the return ionization current. Other circuits are used to provide filtering before being digitized by an ADC. After the conversion, the microcontroller has to compute the reverse transfer function from voltage to pressure.

## 3.4 Hot cathode ionization gauge (Bayard-Alpert)

To measure down to $10^{-12}$ mbar, Bayard-Alpert gauges are used. A heated filament emits electrons that are accelerated by the grid potential (+150 V). On their path, these electrons ionize gas molecules, which are gathered by a collector. The collector current is a measure of the ionization and therefore of the gas pressure:

$$I^+ = S.I_e.p$$

The factor $S$ is the sensitivity, $I_e$ is the emission current and $p$ is the pressure.

The ionization current can be as low as 100 fA and thus, a high quality triaxial cable is used for this measurement. For the measurement of the ionization current, the conditioning circuit is made by an ammeter in a piecewise linear configuration (see chapter 3.5), meaning that the gain changes according to the range of measurement. Several power supplies are used to heat the filament, to measure the emission current, and to bias the filament and the grid. A stepdown voltage transformer is used to supply the filament circuit to overcome the voltage drop due to the resistance of the cable.

## 3.5 Low current measurement

### 3.5.1 Ammeter techniques

In most instrumentation applications today, two common methods of current measurements are available: the shunt ammeter method and the feedback ammeter method [5].

In the shunt ammeter method, low values of shunt resistors are chosen to minimize the voltage dropped across the shunt. Although the voltage drop is small, this can have a negative impact on the circuit under test and the measurement for low currents. This voltage drop is known as the voltage burden and is a series voltage error introduced by an ammeter. If the shunt resistor of the ammeter is too large versus the resistances in the circuit under test, the voltage burden causes large errors.

Feedback ammeters as displayed in Figure 3, on the other hand, use a different method to produce a current measurement. They use an active transimpedance amplifier to convert the current to a voltage reading. The voltage output is the inverse of the current input multiplied by the value of the feedback resistor $R_F$. With the feedback ammeter method, voltage burdens are much lower. Feedback ammeters also have the advantage of measuring low currents much faster.

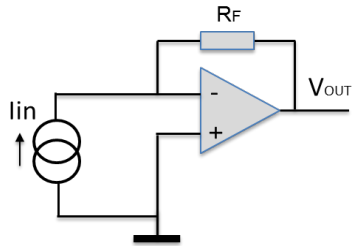

**Figure 3 – Feedback ammeter**

A practical limitation of the feedback ammeter method is its high-current capability. Feedback ammeters are generally designed to measure currents of up to 20 mA. At that level, the shunt ammeter method, in which the shunt resistors are small enough to deliver fairly fast settling times, becomes a better option.

The dynamic range has to be taken into account for vacuum measurements. For example, Penning gauge measurements require 7 decades of pressure (from $10^{-5}$ down to $10^{-11}$ mbar) and current. For that, two main current measurement techniques are used: the piecewise linear amplifier and the logarithmic amplifier.

The piecewise linear amplifier illustrated in Figure 4 uses the feedback ammeter configuration with several discrete gains. It requires the use of a microcontroller to manage the change of gains according to the measurement range and to perform some digital processing.

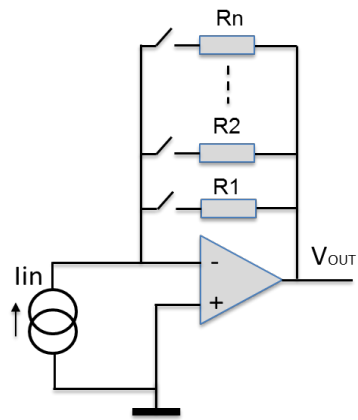

**Figure 4 - Piece-wise linear ammeter**

Its output voltage is the inverse of the input current multiplied by the combination of the feedback resistors. For very high gain, $10^{12}$ ohms resistors are needed and special relays with low leakage current are used.

The logarithmic amplifier displayed in Figure 5 converts a linear measurement into a logarithmic function and thus compress the dynamic range; it is fully analog.

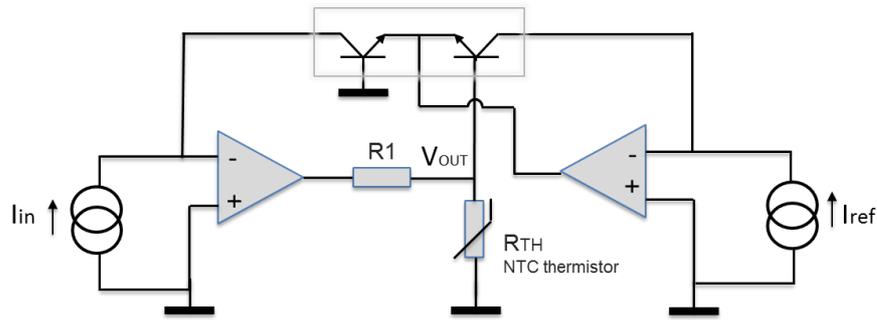

**Figure 5 - Logarithmic ratio ammeter**

Its voltage output is a logarithmic function of the ratio between the input current and a reference current:

$$V_{OUT} = \left(1 + \frac{R_1}{R_{TH}}\right) \cdot \frac{k.T}{q.log_{10}(e)} \cdot log_{10}\left(\frac{I_{IN}}{I_{REF}}\right)$$

The factor $q$ is the electron charge, $k$ the Boltzmann constant and $T$ the absolute temperature. This electrometer is sensitive to temperature but this effect can be slightly compensated by using a NTC (Negative Temperature Coefficient) thermistor $R_{TH}$. It is also important to use a matched pair of transistors (two identical transistors in the same package) so that the transfer function above can be verified.

### 3.5.2 *Interconnections and cabling*

With low-current measurements, interconnections and cabling are of a great importance to ensure maximum noise reduction and optimal shielding. The most common sources of noise and leakage are the following [5]:

- The 50/60 Hz line noise pickup is the most common and considerable source of noise. This effect can be reduced by the usage of proper cabling, including shielded and coaxial cables.

  It must be noted that poor cable shielding cannot be compensated with filtering. The 50/60 Hz line-induced noise can easily saturate the sensitive preamplifier circuitry of any low current ammeter. Once this happens, no amount of filtering can recover measurement accuracy. Therefore, shielded cables are a must.

- Triboelectric effects arise from the movement of a conductor against an insulator. Cables designed to minimize this effect are available. Reducing cable movement with tie-downs or other fixtures also minimizes this effect.

- Piezoelectric effects are caused by the physical deflection of an insulator. Reducing the amount of mechanical stress helps to minimize this effect.

- Leakage currents are often the result of contaminants around the device of interest. These contaminants provide an additional current path that causes measurement errors. Solder flux and fingerprints (oil, salt, etc..) can be sources of contamination. Many contaminants can be cleaned using alcohol or a similar solvent.

# 4 Vacuum pumps and controllers

## 4.1 Ion getter pumps

Sorption in ion getter pumps relies on sputtering of a getter material inside a gas discharge, and additionally, on the bombardment of ions from the gas discharge. Gas discharge in an ion getter pump is of the Penning type. In the range from $10^{-5}$ down to $10^{-9}$ mbar, the ion getter pump can be used to measure the pressure, in addition to its pumping function.

The getter pump is composed of several Penning cells, with Titanium cathodes and a potential of 6 kV between the anode and cathodes. In the low range of measurement, the current can be as low as 100 nA for small pumps, and high quality HV coaxial cables are used for signal transmission.

HV can be produced by using a linear high voltage step-up transformer, with its primary directly connected to the mains, followed by a voltage multiplier. The transformer size will depend on the power required.

Another technique consists of using a switching mode power supply, composed of a high voltage and high frequency step up transformer, with its primary connected to a DC voltage and its secondary connected to a voltage multiplier. The switching frequency at the primary of the transformer is in the order of several decades of kHz, reducing the size and the weight of the power supply. Moreover, the output voltage can be easily adjusted accordingly to the current provided to the pumps. This allows a reduction of the leakage current and optimises the pumping speed.

The controller has to provide a measurement of the ionization current, which is a measure of the pressure as for the Penning gauge (see chapter 3.3).

## 4.2 Sublimation pumps

A titanium filament is supplied with a high current (typically 40A). It is heated until it reaches the sublimation temperature. The surrounding chamber walls are coated with a thin film of clean titanium and act as pump for reactive gases.

The power supply of sublimation pumps provides AC power modulation by using a thyristor. A step-down voltage transformer is used close to the sublimation pump to cope with the length of the cables. The power supply provides current and sublimation measurement. The controller of the power supplies is based on PLC modules integrated in a 3U Europa crate. It can handle several power supplies and is connected to the master PLC.

## 4.3 Roughing and turbo molecular pumps

Turbo molecular vacuum pumping groups are used for rough pumping, for leak detection and to maintain vessels under high vacuum. The primary pump performs initial pumping from atmospheric pressure to rough vacuum and then the turbo molecular pump achieves high vacuum [7].

Figure 6 shows the basic arrangement of a turbo-molecular pumping group. It is composed of a primary pump (VPP), a turbo-molecular pump (VPT), several valves (VVP, VVI, VVT, VVD and VVRs) one or more gauges (VG).

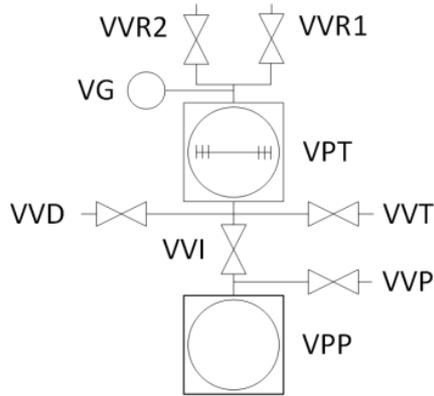

**Figure 6 - Basic structure of a turbo-molecular pumping group**

The management of a pumping group is an automated process handled by a PLC integrated in a 3U Europa crate, which include all of the required hardware and software to independently control the pumping process. Vacuum pumping groups are either fixed (permanently connected to the volumes) or mobile (mounted on wheeled trolleys and temporarily installed wherever they are required).

Mobile pumping groups are connected to the Profibus network available in the accelerator tunnels, allowing the connection to the SCADA, via the master PLC. Figure 7 shows the control architecture for pumping groups [6].

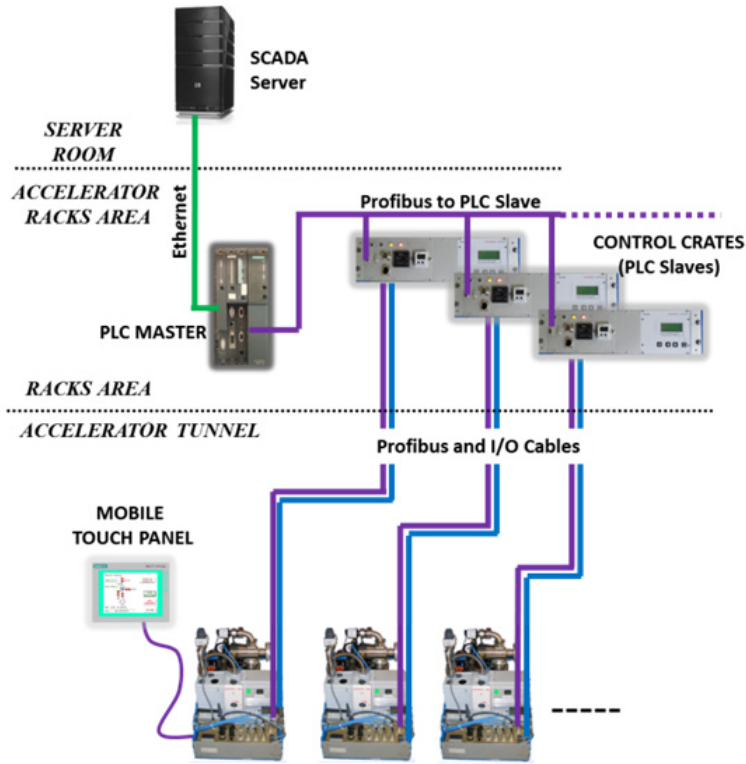

**Figure 7 - Typical control architecture for pumping groups**

The pumping group operation is started by turning the VPP on. Once the primary pump has reached nominal state, either detected by monitoring the pressure between the VPP and the VVI valve, or assumed after a given time, the VPT is started and the VVI is opened.

VPTs rotate rapidly and require a dedicated controller. VPTs reach their nominal state at around 1000 rpm, at which point the VVR valves may be opened to pump the vacuum vessel.

Multiple VVRs may exist in order to pump different vessels. The VVP and VVT valves are used to vent the pumps, while the VVD is used to connect a gas analyzer for leak detection. Pressure at the inlet of the VPT (which is indeed the vessel pressure once VVR is open) is monitored by the VG, which is a full-range set of Pirani/Penning gauges.

### 4.4 Bake-out control

CERN Accelerators have more than 7 km of beam vacuum vessels, which operate at room temperature and require Ultra-High Vacuum. During installation or after a vacuum intervention, the vessels require conditioning; this consists of a heating cycle (bake-out) over several days, with a dual role [8]:

- A fast and large outgassing of the vacuum vessels: after the bake-out, vacuum vessels must have significantly reduced their outgassing rate.
- A thermal activation of the Non-Evaporable Getter (NEG): the oxide layer present at the NEG surface is dissolved, releasing a high pumping speed thin film.

The bake-out cabinet is based on Programmable Logical Controllers (PLC), comprising a CPU, Input/output modules, fieldbus and network interfaces. It is integrated in a standard 13-unit high and 19-inch width rack, mounted on wheels.

Fieldbus networks (Profibus) dedicated to mobile pumping groups and mobile bake-out cabinets are installed in the accelerators tunnel. The bake-out cabinets are connected to the master PLC via the Profibus interface. The master PLC is used as a gateway to the SCADA server. Once connected, the bake-out cycle can be monitored by the vacuum SCADA application. Figure 8 shows the control architecture for bake-out cabinets.

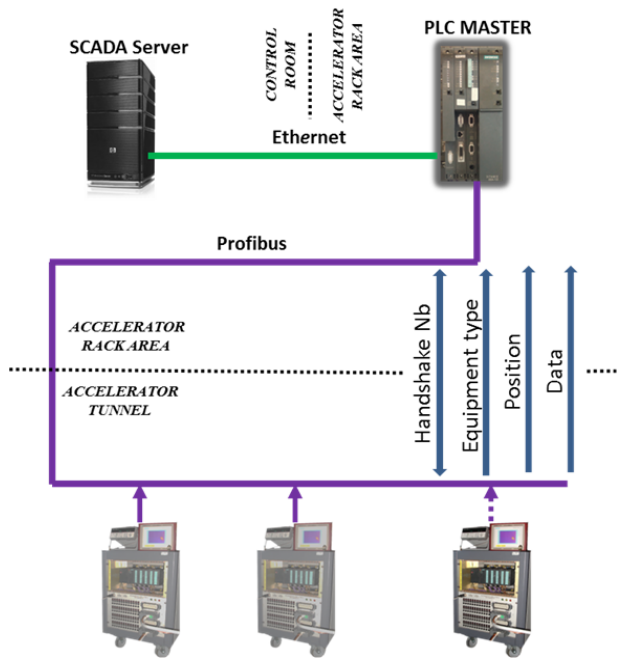

**Figure 8 - Typical control architecture for bake-out cabinets**

# 5 Hardware interlocks and alarms

Alarms are binary signals sent to the operator consoles, e-mail or SMS, to draw their attention to particular equipment status or situations. There are also hardwired alarms connected to other systems, for information or to be used in their control logic [9].

Interlocks are binary signals used within the vacuum system to prevent entering in undesired states. They have a higher priority than the normal process logic or the operator commands. Apart from a beam permit interlock, the vacuum control system does not interlock any other external systems directly.

The detection of a pressure rise, above a predefined threshold, can be used to produce alarms or interlocks. Membrane, Pirani, Penning and Ion Pumps can be used for alarms or interlocks.

## 5.1 Interlocks for the vacuum sector valves

Beam pipes are divided in "vacuum sectors" that can be isolated by "sector valves" (VVS), to avoid the propagation of leaks over a large volume. Upon the detection of a pressure rise above a predefined level, these valves will close in 1 to 3 s, depending on the model.

In general, a sector valve is interlocked by several neighbouring pressure measurements, either to make it close or to disable opening. Pressure thresholds can be set individually for each gauge, but preference is given to uniform values all around the machine. A valve closed due to a pressure interlock will also interlock its 2 neighbour valves, to avoid pressure increase propagation. Figure 9 shows the typical interlock configuration between two vacuum sectors at room temperature.

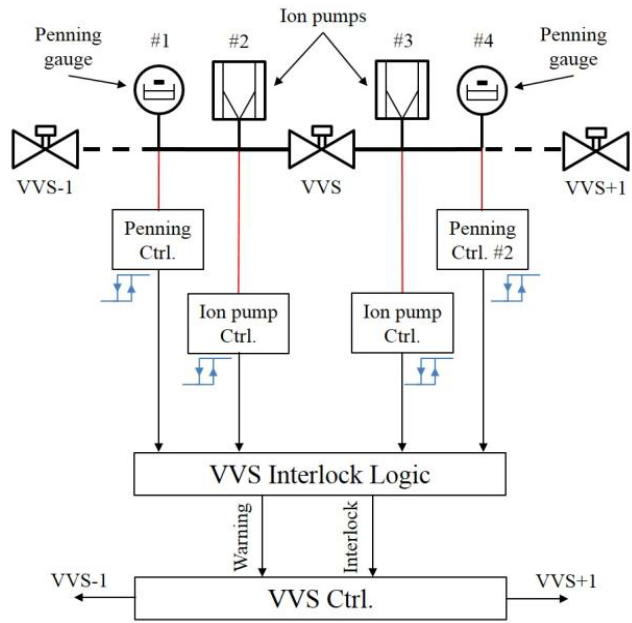

**Figure 9 - Typical interlock configuration for vacuum sector valves in LHC**

## 5.2 Interlocks to the Beam Interlock System (BIS)

As a closed sector valve would be in the beam path, and would consequently get damaged, the beam must be stopped before the valve closure. The beam can be extracted from the LHC and dumped in less than 300 μs, which is much faster than any valve closing (1 - 3 s). Therefore, a pressure interlock triggering a valve closure is simultaneously forwarded to the BIS; and so is the information of any valve losing the open status, due to a neighbour valve closing or for any other reasons [10].

### 5.3 Alarms to the cryogenic system

If the pressure rises, the thermal isolation degrades and the cryogenic system might need to prevent the start of some cryo-compressors or even to stop them. Each insulation volume around the cryogenic distribution line (QRL) or the cryo-magnets is equipped with a triplet of gauges (Membrane, Pirani and Penning). The first two are used to generate alarms to the cryogenic system.

### 5.4 Alarms to the RF & ADT system

If the vacuum degrades inside a radiofrequency (RF) accelerating cavity, the high electrical field may ionize the residual gas and damage the cavity. For each of the 8 cavities in each beamline, the RF system has its own Penning gauges, to follow the pressure inside the cavities, and to produce the necessary interlocks. These are not available on the vacuum system. The pressure in the beam pipes outside of an RF module (group of 4 cavities) is used to produce an alarm, combining one Penning gauge and two ion pumps. For the transverse dumpers (ADT), located upstream and downstream of the modules, the RF system receives the analog reading and a Penning alarm.

### 5.5 Alarms to the kicker systems

The injection kicker magnets (MKI) insert the beam coming from the transfer lines into the LHC orbit; there are 4 MKI tanks at point 2 and 4 other at point 8. The dilution kicker magnets (MKB) sweep the beam in an 'e' shape, distributing it over a wide dump surface; there are 5 MKB tanks on each side of point 6. As the kicker magnets are powered by high-voltage pulses, if vacuum degrades, galvanic isolation becomes a concern and the kickers may have to be stopped. If pressure rises in the MKI, there cannot be injection into the LHC. If pressure rises in the MKB, the beam should be dumped before the situation gets worse.

## 6 Supervision

### 6.1 Introduction

The supervision layer of control systems has radically evolved over the last decades from point-to-point wiring, where control rooms were mostly based on analog gauges and indicator lights directly connected to field equipment, to computer-based networked solutions. As fieldbuses, network protocols and computing power evolved, modern control rooms are now practically free of point-to-point wiring, where the latter are usually reserved for mission-critical applications. Although this allowed for a significant decrease of cost and complexity with cabling, the complexity of software applications responsible for supervision has increased dramatically.

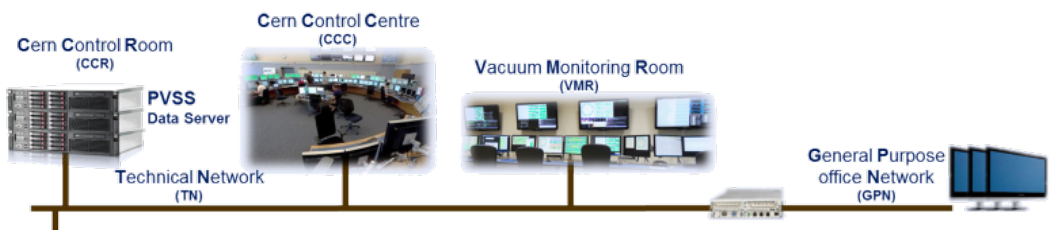

**Figure 10 – The Supervision Layer**

On the supervision layer, a data server communicates with the devices on the control layer to create an image of the underlying control processes and equipment, so it can be shared with the users through dedicated consoles, and receives commands from these consoles and propagates them to the respective equipment or process. The software responsible for this task is commonly known as SCADA, acronym for Supervisory Control And Data Acquisition. The SCADA software typically provides:
- User Interfaces (UIs) so that operators can monitor and control processes and equipment
- Alarm handling features
- Historian functionalities such as data logging and retrieval through historical panels and trends
- Access control mechanisms
- Offline data processing

Many solutions are available, proprietary and Open Source, being the most common within the particle physics community WinCC-OA, TANGO and EPICS. Independently of the solution chosen, the key aspects to be taken into consideration when choosing a SCADA solution are openness, scalability, connectivity and multiplatform ability.

Openness refers to the easiness in expanding core functionalities so that custom features can be developed. Out of these, commonly required features are integration (data-sharing) with external clients, process-specific data analysis and alarm logic, or high-level control logic.

Scalability deals with the ability for the system to cope with the increase on the number of controlled devices, datapoint elements, users, consoles, and communications load.

Connectivity refers to the availability of communication drivers supported by the devices in the control layer, and whether these protocols and their underlying networking infrastructure are able to provide adequate levels of latency.

Finally, in an organization where several operating systems might coexist, a point to consider might be the ability of the SCADA system to work in multiple operating system environments (Windows, Linux, macOS).

Because it met all of the above-mentioned criteria on the specific context of CERN, WinCC-OA (open architecture) was selected to be the SCADA solution for all vacuum applications, being also used in the majority of all accelerator systems and its supporting infrastructures.

## 6.2 WinCC-OA architecture

WinCC-OA is built using a modular architecture where each functionality is handled by a specific unit, called a manager. It is composed of 4 layers as illustrated in Figure 11**Error! Reference source not found.**. Starting from the bottom, we have:
- Driver managers: responsible for handling the communication between the supervisory application and the devices in the control layer (Ex: Siemens S7, Modbus TCP/IP)
- Process Image and History: provide current and past state of system variables (Ex: Event manager, database manager)
- Control and API layer: composed of scripts (using a dedicated language called CTRL) providing customized behavior and application programmer interfaces (APIs). Examples of scripts in this layer include SMS alert features based on pressure values, valve statuses, etc.
- User interfaces: graphical panels that display process data and allow the interaction between operators and the process

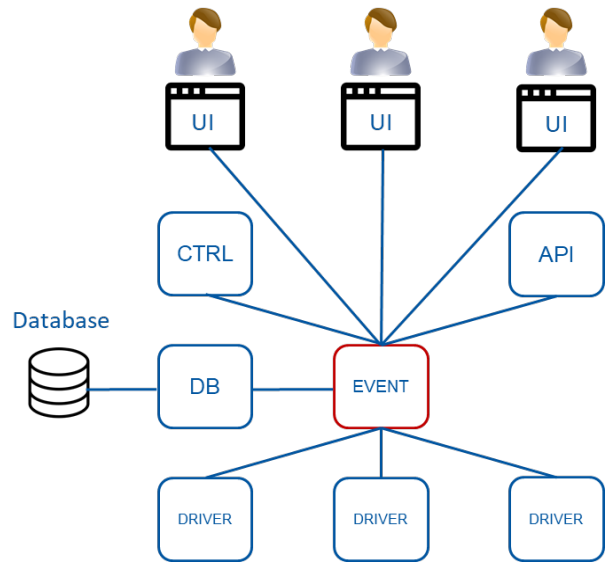

**Figure 11 – WinCC-OA Managers Architecture**

It can be seen that the architecture is built around the event manager, acting as the central hub of data on WinCC-OA SCADA applications. The event manager receives data from the drivers (and eventually from other managers), and distributes it to the managers who are subscribing to specific sets of data. An important subscriber of this data is the database manager, responsible for the archiving and retrieval of historical data, being a fundamental element for features such as alarm handling and trending. Another key feature of WinCC-OA worth mentioning is the fact that managers can be organized to run in a distributed fashion by running on different machines. In this way, the processing power required to run a SCADA application can be distributed amongst many machines. As an example, intensive data analytics CTRL scripts can be made to run on a different machine to minimize any impact on performance on the machine running the event and driver managers.

### 6.3 Data Flow

The structure and key elements of data flow in vacuum controls applications at CERN is illustrated in Figure 12. Every PLC contains 2 dedicated Data Blocks for communication with the SCADA, one called the Read Register (RR) used by the SCADA to read PLC data, the other called the Write Register (WR) used by the SCADA to send data (typically commands or parametrization) to the PLC. These communication Data Blocks are large buffers in the PLC memory, where data of all devices is stored in positions specified in the device Data Block instances. An overview of the memory mapping between the SCADA and PLCs is illustrated in Figure 12.

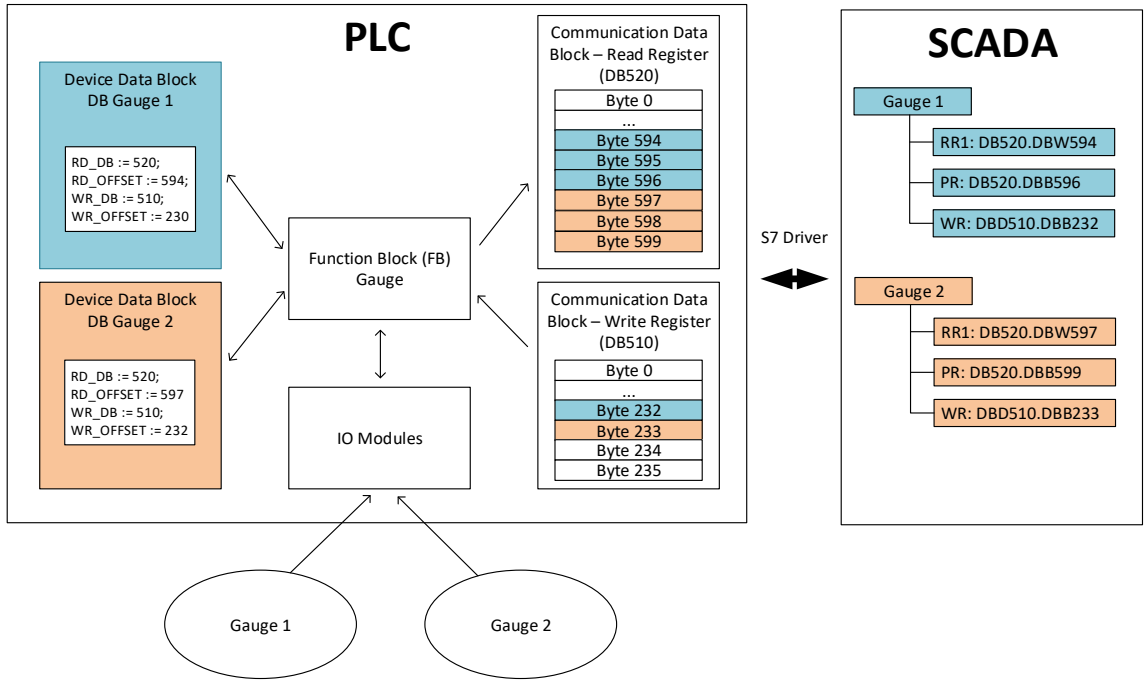

**Figure 12 - Memory mapping**

In order to make device data available to the SCADA, the PLC reads the physical state of the field devices from its input/output (IO) modules. After processing this raw data, device Function Blocks copy the relevant data to the PLC Read Register based on configuration defined in the device Data Blocks. In Figure 12, as an example, the instance Data Block of Gauge 1 (in blue) contains configuration that will instruct the gauge Function Block that its data is to be written to DB520 (RD_DB := 520), on offset 594 (RD_OFFSET := 594). Similarly, Gauge 2 (in orange) will be written to the same Datablock, with a different offset (597). In general, all devices controlled by the PLC will have assigned a unique memory location within the PLC read register. On the SCADA, devices are declared as datapoints (ex: Gauge1, Gauge2), each datapoint being composed of datapoint elements (ex: RR1 containing the status of the device, PR containing the pressure value) whose S7 address configuration mirrors the memory locations on the PLC read register. The S7 driver on the SCADA will poll the PLC at fixed rate in order to refresh its image of the PLC memory.

For data transfer between the SCADA and the PLC, device Write Registers are modified in the SCADA by either User Interfaces (when for instance a user presses a button to switch on/off a gauge), or CTRL scripts, and the S7 driver will transfer their value to the appropriate location in the PLC memory. The Function Blocks responsible for handling the equipment type on the PLC will act according to the data received from the SCADA.

## 6.4 Data Engineering

In the previous section, it was seen that in order for the communication between the SCADA and PLCs to work, there must be a mapping between memory locations in the PLC and in the SCADA. Although this mapping can be manually configured for small control systems, in large applications containing thousands of devices with potentially hundreds of thousands of datapoint elements, it is simply unpractical, if not impossible, to handle configuration without specific tools. Furthermore, besides the handling of memory locations, other functionalities of the control system need to be configured: these include alarms, short- and long-term archiving, the publishing of vacuum data to other control systems, and the subscription of data from other control systems that might be of interest for the vacuum control system (namely cryogenics and accelerator access control). Figure 13 illustrates an overview of the various elements involved in the data engineering process of the vacuum control system at CERN.

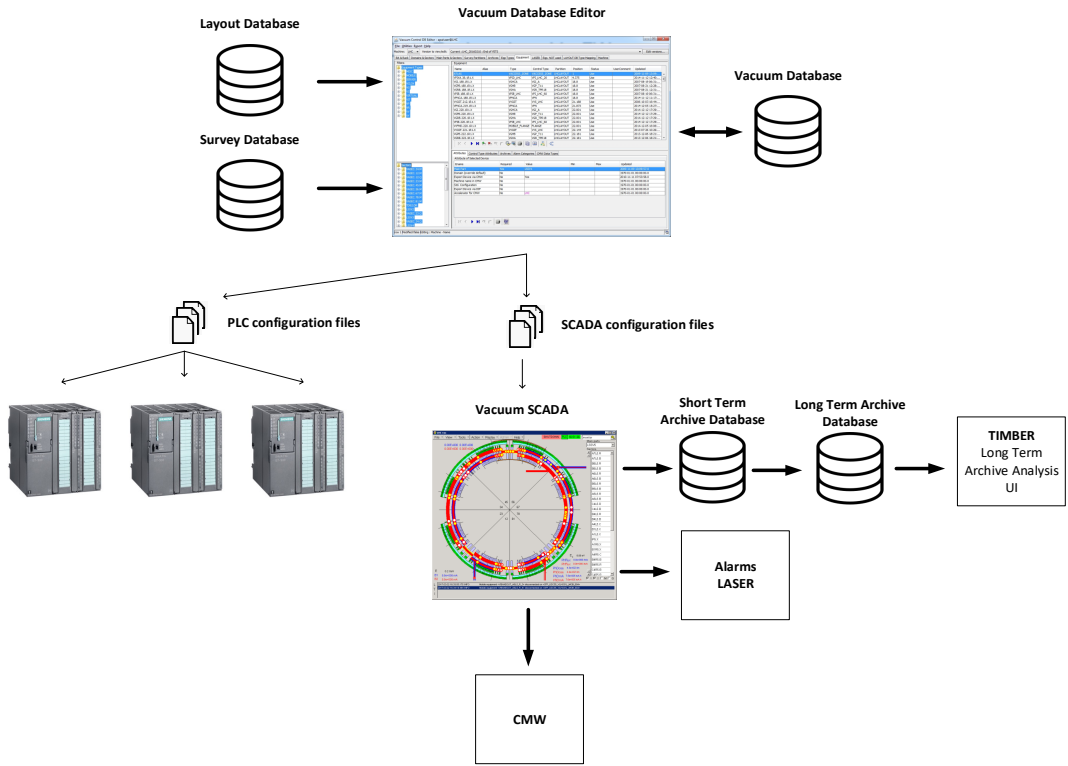

**Figure 13- Overview of data engineering scope**

The configurations mentioned above are stored in a set of Oracle Databases, called the Vacuum Database. From the data in Vacuum Database, an export module calculates memory locations and generates automatically the configuration files for the SCADA and PLCs, mapping SCADA datapoints to memory locations in the PLC. The configuration files for the SCADA include data on which datapoint elements are to be published to CERN's central alarm system, called LASER (LHC Alarm Service), short- and long-term archive configurations defining which datapoint elements will be archived with their respective deadbands, and which vacuum values should be published in CERN's CMW (**C**ontrols **M**iddle**w**are**)** infrastructure to allow other control systems to access vacuum data. In addition to configurations stored in the Vacuum Database, the Vacuum Database Editor uses configuration data from 2 databases external to the vacuum service. The first is CERN's Layout Database, that contains equipment names along with their structure and position along the accelerator, allowing the vacuum SCADA to place equipment in the appropriate vacuum sectors in order. The second database is the Survey Database, a three-dimensional database that contains geographical positions of accelerator equipment, allowing the SCADA to generate a special representation of the accelerator systems.
The SCADA configuration files are then imported into the SCADA via a dedicated import panel, whereas PLC sources (part of the PLC configuration files) are downloaded into the PLCs.

## 6.5 Vacuum SCADA functionalities

While the previous sections describe concepts generally applicable to any control system, this section focuses on some SCADA functionalities that are of particular interest for vacuum applications.

### *6.5.1 Main Views*

The main view provides a general view of the layout of the accelerator state with beam lines coloured according to the state of sectors valves. The layout is automatically built with information taken from the Survey Database (see Figure 14). Zones of the beam line where sector valves are opened, allowing the beam to pass, are coloured in green, whereas zones that have valves closed are represented in red. Small discs, color-coded according to the interlock condition active, represent interlock conditions on sector valves.

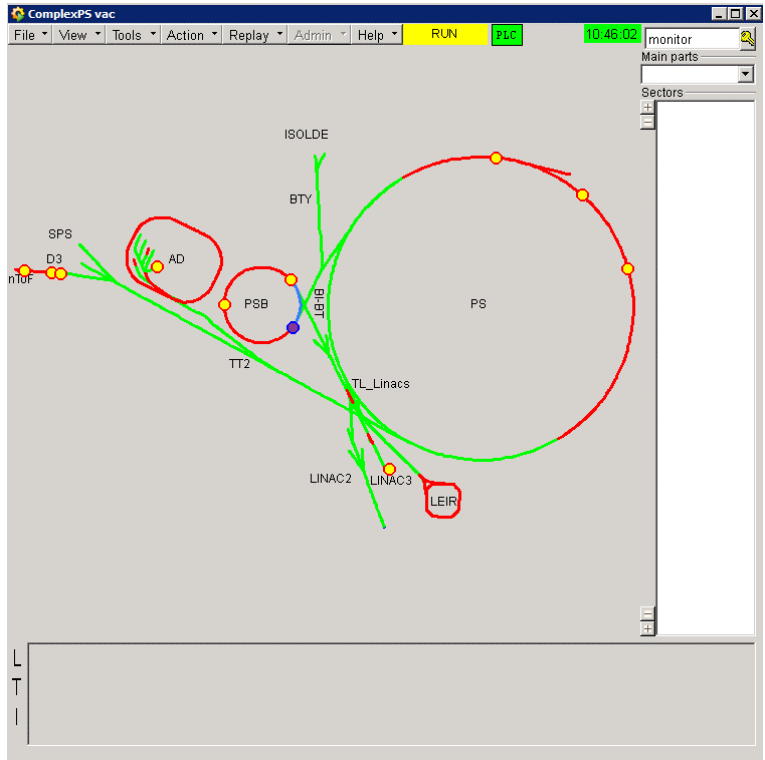

**Figure 14 - Main View**

### *6.5.2 Synoptic View*

The synoptic view represents vacuum devices in order according to their accelerator longitudinal position in their respective beam pipes, as illustrated in Figure 15, where four parallel beam pipes are displayed (LHC blue and read beam pipes, cryogenics distribution line, and insulation vacuum for magnets). Each type of device has a particular shape and follows a colour convention, allowing operators to easily identify the type and status of equipment. Extra information, such as magnet and sector names, are placed on the synoptic to aid operators in locating vacuum equipment by providing references.

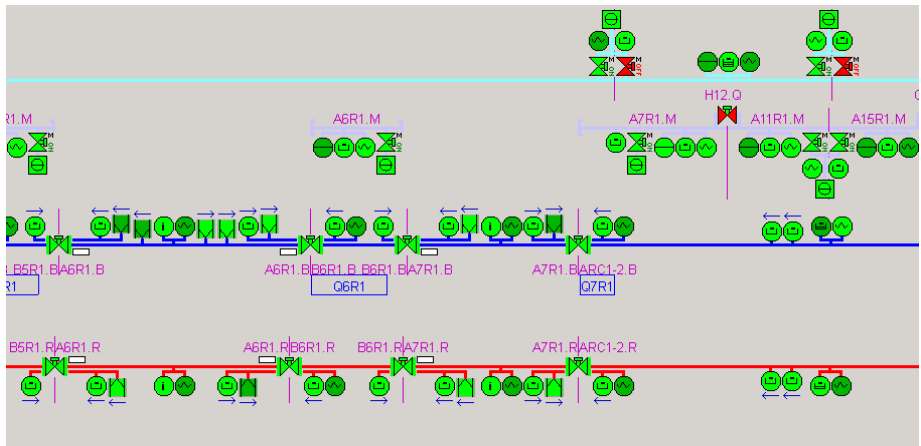

**Figure 15 - Synoptic View**

### 6.5.3 *Pressure Profile*

The pressure profile displays pressure along the longitudinal coordinates of the accelerator on bar graph, allowing operators to easily identify overpressure situations, and zones that might contain a vacuum leak. In addition to displaying pressure, the pressure bars, read by individual gauges, are colour coded according to their status, so that faulty or off gauges can be easily identified.

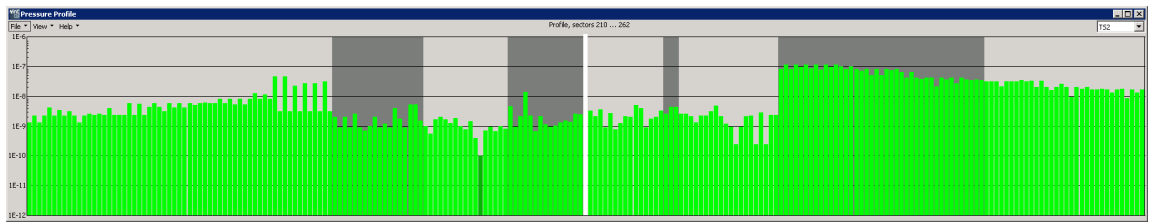

**Figure 16 - Pressure Profile**

### 6.5.4 *Equipment Details Panel*

Although it is possible to have an overview of the status of equipment through the synoptic view, equipment detail panels offer more detail on the status of equipment and allow authorized users to issue commands. Figure 17 illustrates an example of a detailed panel of a pumping group.

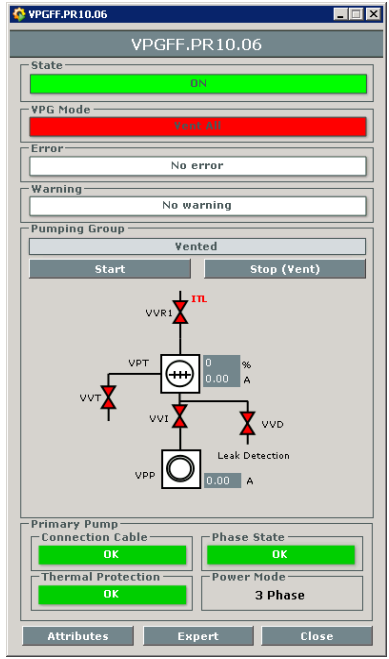

Figure 17 - Example of equipment details panel for a pumping group

### 6.5.5 Multiline Trend

The multiline trend, displayed in Figure 18, is a diagnostic tool that allows to plot several measurement graphs in the same panel, allowing operators to correlate different measurements. As an example, Figure 18 displays a multiline trend panel with pressure, temperature, and beam energy. Different measurements are displayed with different colours, and scales for each unit can be adjusted individually. The panel displays online data as well as historical data.

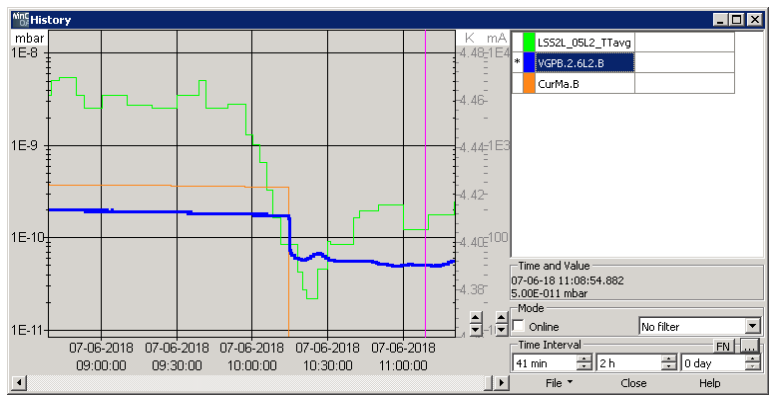

Figure 18 - Multiline Trend

### 6.5.6 Equipment State History

The equipment state history displays historical data of the status of equipment. Each event that lead to a change of status of a device is displayed in a new line, with a status text and colour that represents the state of the device. In order to help in diagnosing controls related problems, for each event, the bottom part of the panel displays the status of all bits on the read register (see Section **Error! Reference source not found.** for details on the read registers). Several devices can be displayed in the same window, enabling the correlation of events.

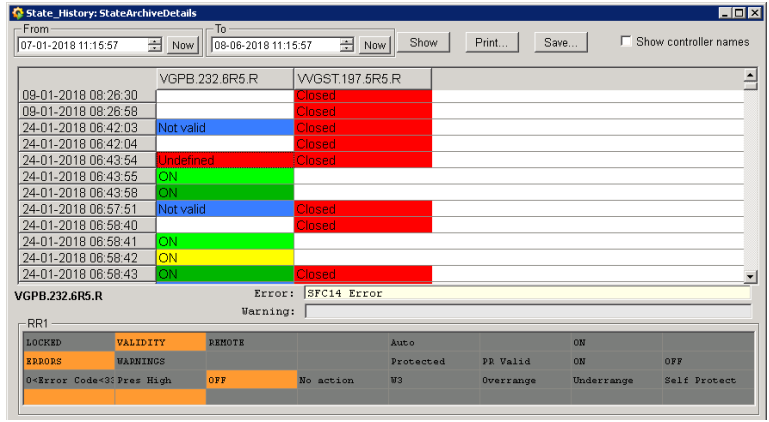

Figure 19 - State History

### 6.5.7 Equipment Help Panels

As the number of equipment functionalities and types increase, it becomes more and more difficult for operators to understand the meaning of vacuum equipment widgets. Equipment help panels are accessible from the synoptic view, and provide operators with an interpretation of the meaning of all shapes and colours seen in the device representation. An example of a sector valve help panel is illustrated in Figure 20.

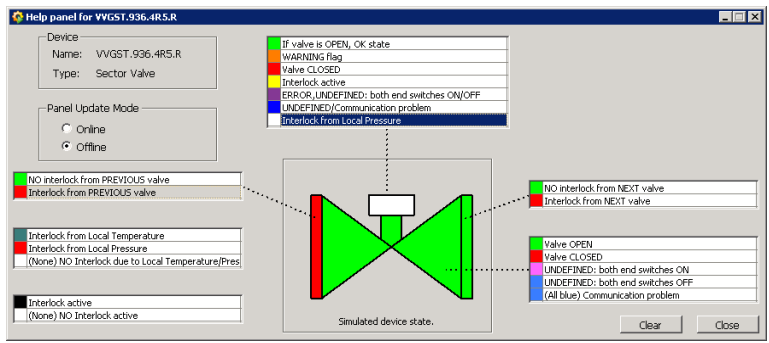

Figure 20 - Sector Valve Help Panel

### 6.5.8 Sharing Vacuum Data

The vacuum control system is merely a subsystem of a global accelerator control system. While it can be independently controlled, accelerators have to be managed as a whole, and for this reason several data sharing mechanisms are in place. Figure 21 illustrates several examples of systems interacting with the vacuum SCADA at CERN.

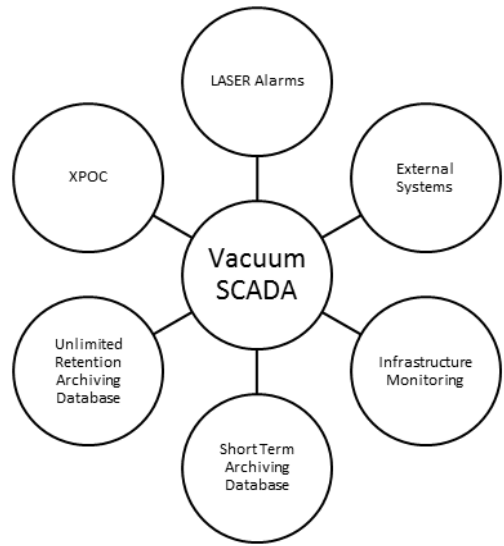

**Figure 21 - Systems interacting with CERN's vacuum control system**

### 6.5.9 LASER

Vacuum alarms are published to a central alarm system, called the **L**hc **A**larm **SER**vice, that agglomerates alarms from the various accelerator subsystems at CERN. The configuration for the LASER alarms is automatically generated from the Vacuum Database Editor, and is parsed by a dedicated manager in WinCC-OA that is responsible for detecting changes in data point elements and publishing them to LASER in case alarm conditions are detected. More information on LASER can be found in [11].

### 6.5.10 External Systems

The vacuum SCADA interacts with the following accelerator subsystems:

- Beam Operations: so that the vacuum control system is aware of the current accelerator mode (RUN, STOP). Several SCADA functionalities depend on this feature. As an example, a closed sector valve is considered an alarm condition when the accelerator is in RUN mode, but not when the accelerator in STOP mode.
Furthermore, when access is given to a particular zone in the accelerator, the accelerator sequencer issues commands to the vacuum control system to close the sector valves in that zone, for safety reasons. The vacuum control system also receives information concerning beam dump events, useful for the functionality described in section 6.5.14.
- Beam Instrumentation: so that beam data, such as energy, momentum and filling schema can be represented in the vacuum SCADA to be correlated with vacuum data.
- Cryogenics control system: the vacuum control system sends pressure data to the cryogenics control system, and receives from it values of temperatures in magnets.

### *6.5.11 Infrastructure Monitoring*

The status of vacuum data servers, containing available memory, disk space, and CPU utilization, along with the status of WinCC-OA managers and PLC statuses are published to a central service called MOON, where the infrastructure of all vacuum applications can be visualized through a web application. Application responsibles are notified when infrastructure related alarm conditions are met.

### *6.5.12 Short Term Archiving Database*

The short-term archiving database, commonly known as RDB, is responsible for archiving vacuum data. A WinCC-OA manager, called the RDB manager, is responsible for sending data to this database, taking into consideration archive smoothing configurations defined in the Vacuum Database Editor. The RDB database is the source of data for several functionalities mentioned in previous sections such as the trend and state history panels. The limited amount of data, both in archive depth of 2 years and in the number of measurements logged as a function of the defined smoothing parameters, ensures that the data is retrieved from the SCADA with an acceptable latency.

### *6.5.13 Unlimited Retention Archiving Database*

While the short-term archiving database mentioned in the previous section allows fast access to historical data directly on the vacuum SCADA, an unlimited retention archiving database, known as the LHC Logging DB holds data from all accelerator subsystems without limits on archive depth. Due to the significant amount of data that is contains, data retrieval is considerably slower when compared to the short-term archiving database. The access to this data is done using a dedicated application, called Timber.

### *6.5.14 External Post-Operational Checks (XPOC)*

After every LHC beam dump, the eXternal Post-Operational Check system automatically verifies the correct execution of the beam dumping control systems and beam instrumentation measurements, to check the correct execution of the dump action and the integrity of the related equipment [12]. When a beam dump event is received by the vacuum SCADA, via CMW, a dedicated manager gathers pressure data at fixed intervals, before and after the dump event, and sends it to the XPOC system so that the analysis can take place.

**Acknowledgement**

We wish to thank CERN TE/VSC Colleagues for their direct and indirect contributions to this paper.